\documentclass[twocolumn,tighten]{aastex63}
\usepackage{natbib}
\usepackage{needspace}
\usepackage{tabularx}
\usepackage{booktabs}
\usepackage{multirow}
\usepackage{amsmath}
\usepackage{tablefootnote}
\usepackage{soul}
\let\tablenum\relax 
\usepackage{siunitx}

\newcommand{\Gaia}{{\textit{Gaia} DR3}}

\newcommand\erosita{eROSITA}

\shortauthors{Aldarondo Quiñones et al.}
\begin{document}
\title{Reassessing the Relationship Between Stellar X-ray Luminosity and Age with eROSITA~Data~Release~1}

\author[0000-0003-1329-176X]{Nadja Aldarondo Quiñones}
\email{nadja.aldarondo@upr.edu}
\affiliation{Department of Astronomy,  University of Wisconsin--Madison, 475 N.~Charter St., Madison, WI 53706, USA}
\affiliation{Department of Physics, University of Puerto Rico-Río Piedras, Facultad de Ciencias Naturales, Edificio Fase II, 17 Ave. Universidad STE 1701, San Juan, PR 00925-2537}

\author[0000-0001-9827-1463]{Sydney Jenkins}
\altaffiliation{National Science Foundation Graduate Research Fellow}
\affiliation{Department of Physics and Kavli Institute for Astrophysics and Space Research, Massachusetts Institute of Technology, Cambridge, MA 02139, USA}

\author[0000-0001-7246-5438]{Andrew Vanderburg}
\altaffiliation{Alfred P.~Sloan Research Fellow}
\affiliation{Center for Astrophysics | Harvard and Smithsonian, 60 Garden Street, Cambridge, MA 02138, USA}
\affiliation{Department of Physics and Kavli Institute for Astrophysics and Space Research, Massachusetts Institute of Technology, Cambridge, MA 02139, USA}

\author[0000-0001-7493-7419]{Melinda Soares-Furtado}
\altaffiliation{NASA Hubble Postdoctoral Fellow}
\affiliation{Department of Physics and Kavli Institute for Astrophysics and Space Research, Massachusetts Institute of Technology, Cambridge, MA 02139, USA}
\affiliation{Department of Astronomy,  University of Wisconsin--Madison, 475 N.~Charter St., Madison, WI 53706, USA}

\author[0000-0001-5226-8349]{Michael A. McDonald}

\begin{abstract}
Accurate stellar dating provides crucial information about the formation and development of planetary systems. Existing age-dating techniques are limited in terms of both the spectral type and age range they can accurately probe, and many are unreliable for stars older than 1 Gyr. Recent studies have suggested that a star's X-ray luminosity correlates strongly with stellar age and shows a steep fall-off at ages older than 1 Gyr. In this work, we present X-ray luminosity relationship values from eROSITA for four previously unassessed stars. Additionally, we reassess the X-ray luminosity/age relationship present in 24 main-sequence stars older than a gigayear. We confirm that a correlation does appear to exist between stellar age and X-ray luminosity at ages older than 1 Gyr. However, we measure a shallower slope with age than previous research for older stars, similar to what was found for younger stars. We also find evidence for significant astrophysical variability in a star's X-ray luminosity, which will likely limit the precision with which X-ray measurements can yield age estimates. We also find weak evidence for mass dependence in the X-ray luminosity/age relationship. These results suggest that although X-ray luminosity correlates with stellar age, it may not serve as a reliable standalone age indicator and is better used as part of a broader suite of age-dating methods.

\end{abstract}


\section{Introduction}\label{sec:intro}


Constraining the ages of planetary systems can provide critical insight into how they form and evolve over time. Given that planet formation occurs on 1-10 Myr timescales, the age of a star can be used as a proxy for the ages of any planets in its system, with most known planets orbiting main-sequence stars \citep{kasting1993habitable}. For instance, stellar ages have been used to probe the formation of Hot Jupiter planets, showing that Hot Jupiters tend to form around younger stars more frequently \citep{dawson2018origins, hamer2019hot, chen2023evolution}. In addition to helping constrain models of planetary formation and evolution, stellar age is also essential for characterizing potentially habitable worlds. It provides an upper bound on how long life has had to emerge, with current research placing the habitability boundary---the earliest point at which Earth became habitable---between 4.5 and 3.9 Ga \citep{pearce2018constraining}. 
Age constraints will be particularly crucial for the upcoming Habitable Worlds Observatory (HWO), which has a primary objective of identifying at least 25 potentially habitable worlds \citep{mamajek2023nasa}.  
Despite the importance of determining the ages of main-sequence stars, current techniques are limited. 
For instance, isochrone fitting and lithium depletion boundaries are most precise when applied to groups of coeval stars (e.g. clusters, \citealt{soderblom2010ages}), which limits usage on a wider range of stars. Additionally, if the planet-hosting star has a white dwarf binary companion, then the age of white dwarf can be used to constrain the age of the system as a whole \citep[e.g.,][]{kiman2022wdwarfdate}. However, this approach is limited to only a small sample of systems, given the relative rarity of white dwarf companions. A more versatile age-dating technique is asteroseismology, which uses the different oscillation modes present in stars to precisely determine the age of stars \citep{kurtz2022asteroseismology}. Advancements in asteroseismological age dating, specifically with data from the Transiting Exoplanet Sky Survey (TESS; \citealt{Ricker2015JATIS}) and ground-based spectrographs like ESPRESSO \citep{Pepe2021A&A} and KPF \citep{Gibson2016SPIE}, provide an extensive database of solar-like oscillators \citep[e.g.][]{hatt2023catalogue, Campante2024A&A,Hon2024ApJ} which can be dated using asteroseismological measures. However, this method requires precise observations \citep{kurtz2022asteroseismology}, limiting the number of stars that can be dated.


Gyrochronology, the study of the rotational spin-down experienced by stars as they age, has previously been used as a way to measure the ages of stars \citep{barnes2003rotational}. However, rotation periods are challenging to measure for older stars, and the relations are not well calibrated for stars older than a few Gyr, limiting the applicability of this technique. Apart from using a star's rotation, there have also been efforts to use the reduction in magnetic activity as an age indicator. The first attempts at defining magnetic activity with a star's age used Ca II emission \citep{Sun_2023, noyes1984rotation, mamajek2008improved}, average equatorial velocity, and lithium abundance. The emission decay was found to be inversely proportional to age. Subsequent research has found variations in the shape of the Ca II profile, which complicate the development of an age-activity relationship. 

Stellar X-ray luminosity offers a possible alternative method for measuring stellar ages. Like Ca II emission, X-ray activity can be used as a tracer of stellar magnetic activity and therefore age. Using known asteroseismological ages, it is possible to define an age-activity between a star's X-ray luminosity and age. 
As previous studies have determined, there is a saturation of the X-ray luminosity until approximately 100 Myr when the X-ray luminosity starts to decay \citep{pizzolato2003stellar, vilhu1984nature}. \citet{jackson2012coronal} investigated the age-activity relationship using cluster data and found slope values ranging from $-1.09 \pm 0.28$ to $-1.40 \pm 0.11$ for  F- to M-type stars younger than 1 Gyr.

\citet{booth2017improved} built upon this work using a sample of 24 stars with well-determined ages older than 1 Gyr. They found that the X-ray luminosity/age relationship has a steep power-law relationship (with slope $-2.80\pm 0.72$) for old ($>$ 1 Gyr) stars, as compared to a shallower relationship (with $-1.09 \pm 0.28$ to $-1.40 \pm 0.11$) for younger stars \citet{jackson2012coronal}. Their findings suggest a steepening of the age-activity relationship that could make it easier to precisely determine stellar age, specifically for stars older than 1 Gyr, from X-ray observations. However, their sample was small (14 stars), so additional stars with both X-ray luminosity and age estimates would be extremely useful to help confirm the trend and refine the relationship. 

Recent date releases provide an opportunity to do just that. NASA's TESS mission has provided seismology measurements for additional stars \citep{nielsen2020tess, chontos2021tess, hatt2023catalogue, metcalfe2023asteroseismology, huber202220}. Meanwhile, the recent release of X-ray data from the Western Galactic Hemisphere from the extended ROentgen Survey with an Imaging Telescope Array \citep[eROSITA, ][]{merloni2024srg} mission has substantially increased the amount and sensitivity of available X-ray data, opening up new windows into stellar activity \citep{zhang2025constraining, zhu2025x, rowan2024precise}. 

In this paper, we probe the X-ray luminosity/age relationship by cross-matching eROSITA X-ray candidates with asteroseismologically age-dated stars. We identify four new sources with both asteroseismic ages and X-ray luminosities and provide updated constraints on the X-ray luminosity/age relationship.  We find that, while X-ray observations remain a possible way to estimate ages, there may be complications due to astrophysical scatter and a shallower slope of the power law.  Our paper is organized as follows: Section~\ref{sec:data} describes the data used in our investigation. Section~\ref{sec:analysis} then describes our data analysis, where we begin by reproducing the results presented by \citep{booth2017improved} before including additional sources from the recent eROSITA release. We discuss the implications of our findings in Section~\ref{sec:discussion} and conclude in Section~\ref{sec:summary}.







\section{Method}\label{sec:data}
To probe the relationship between X-ray luminosity and stellar age, we use a sample of 28 stars with known ages, including 24 stars from \citet{booth2017improved}. Additional X-ray luminosity measurements are included for six of these stars. We also identify 4 new stars with age and X-ray luminosity estimates which we include for the first time here. In this section, we describe our stellar age and X-ray luminosity measurements. Our final sample of stars is given in Table~\ref{tab:sample}, along with all corresponding measurements. 

\begin{table*}
\fontsize{7}{10}\selectfont
\centering
\caption{Full Sample with ages and X-ray luminosities}
\label{tab:sample}
\begin{tabular}{llccccc}
\hline
\hline
\colhead{Name} & \colhead{TIC ID} & \colhead{Age (Gyr)} & \colhead{$\log{L_X}$ } & \colhead{$\log{L_X/R^2}$ } & \colhead{$\log{L_X}$ } & \colhead{$\log{L_X/R^2}$ } \\

& & & \citet{booth2017improved} $^\dagger$ & \citet{booth2017improved} $^\dagger$ & \citet{merloni2024srg}$^\star$ & \citet{merloni2024srg}$^\star$ \\
\hline
alf Men & TIC 141810080 & $6.2^{+1.52}_{-1.52}$ & \nodata & \nodata & $26.89^{+0.06}_{-0.06}$ & $26.92^{+0.06}_{-0.06}$ \\
zet Tuc & TIC 425935521 & $5.9^{+1.17}_{-1.17}$ & \nodata & \nodata & $27.04^{+0.1}_{-0.1}$ & $26.99^{+0.1}_{-0.1}$ \\
pi. Men & TIC 261136679 & $3.8^{+0.9}_{-0.9}$ & \nodata & \nodata & $26.76^{+0.2}_{-0.2}$ & $26.65^{+0.2}_{-0.2}$ \\
88 Leo & TIC 219030951 & $2.4^{+0.4}_{-0.4}$ & \nodata & \nodata & $28.02^{+0.1}_{-0.1}$ & $27.91^{+0.1}_{-0.1}$ \\
16 Cyg A & TIC 27533341 & $6.67^{+0.77}_{-0.81}$ & $26.89^{+0.1}_{-0.1}$ & $26.71^{+0.1}_{-0.1}$ & \nodata & \nodata \\
16 Cyg B & TIC 27533327 & $7.39^{+0.91}_{-0.89}$ & $<25.85$ & $<25.77$ & \nodata & \nodata \\
40 Eri A & TIC 67772871 & $3.7^{+1.34}_{-3.57}$ & $26.81^{+0.1}_{-0.1}$ & $26.97^{+0.1}_{-0.1}$ & $27.07^{+0.07}_{-0.07}$ & $27.23^{+0.07}_{-0.07}$ \\
61 Cyg A & TIC 165602000 & $6^{+1}_{-1}$ & $27.08^{+0.23}_{-0.23}$ & $27.43^{+0.23}_{-0.23}$ & \nodata & \nodata \\
61 Cyg B & TIC 165602023 & $6^{+1}_{-1}$ & $26.88^{+0.1}_{-0.1}$ & $27.33^{+0.1}_{-0.1}$ & \nodata & \nodata \\
CD-3710500 & TIC 179348425 & $1.77^{+0.27}_{-0.65}$ & $28.18^{+0.1}_{-0.1}$ & $28.22^{+0.1}_{-0.1}$ & $28.20^{+0.05}_{-0.05}$ & $28.23^{+0.05}_{-0.05}$ \\
GJ 176 & TIC 397354290 & $2^{+0.8}_{-0.8}$ & $27.03^{+0.1}_{-0.1}$ & $27.8^{+0.1}_{-0.1}$ & $26.72^{+0.17}_{-0.17}$ & $27.49^{+0.17}_{-0.17}$ \\
GJ 191 & TIC 200385493 & $11^{+1}_{-1}$ & $26.41^{+0.1}_{-0.1}$ & $27.02^{+0.1}_{-0.1}$ & $25.73^{+0.15}_{-0.15}$ & $26.35^{+0.15}_{-0.15}$ \\
HR 7703 & TIC 389198736 & $10^{+2}_{-2}$ & $26.8^{+0.1}_{-0.1}$ & $27.04^{+0.1}_{-0.1}$ & \nodata & \nodata \\
KIC 10016239 & TIC 270696968 & $2.47^{+0.61}_{-0.67}$ & $27.91^{+0.1}_{-0.1}$ & $27.71^{+0.1}_{-0.1}$ & \nodata & \nodata \\
KIC 12011630 & TIC 290033019 & $8.48^{+1.42}_{-1.43}$ & $<30.06$ & $<30.05$ & \nodata & \nodata \\
KIC 3123191 & TIC 137892172 & $4.26^{+0.75}_{-0.8}$ & $<29.12$ & $<28.84$ & \nodata & \nodata \\
KIC 5309966 & TIC 172768987 & $3.51^{+0.42}_{-1.23}$ & $<29.44$ & $<29.02$ & \nodata & \nodata \\
KIC 6116048 & TIC 122066308 & $9.58^{+1.9}_{-2.16}$ & $<26.89$ & $<26.74$ & \nodata & \nodata \\
KIC 6603624 & TIC 159574506 & $7.82^{+0.86}_{-0.94}$ & $<27.08$ & $<26.96$ & \nodata & \nodata \\
KIC 7529180 & TIC 63211711 & $1.93^{+0.3}_{-0.35}$ & $28.98^{+0.1}_{-0.1}$ & $28.66^{+0.1}_{-0.1}$ & \nodata & \nodata \\
KIC 8292840 & TIC 159046962 & $3.85^{+0.75}_{-0.81}$ & $<28.14$ & $<27.88$ & \nodata & \nodata \\
KIC 9025370 & TIC 275490621 & $6.55^{+1.13}_{-1.26}$ & $<27.24$ & $<27.23$ & \nodata & \nodata \\
KIC 9410862 & TIC 271042255 & $6.93^{+1.33}_{-1.49}$ & $<27.98$ & $<27.86$ & \nodata & \nodata \\
KIC 9955598 & TIC 270619260 & $6.98^{+0.5}_{-0.4}$ & $26.9^{+0.1}_{-0.1}$ & $27.01^{+0.1}_{-0.1}$ & \nodata & \nodata \\
NLTT 7887 & TIC 129904995 & $4.97^{+3}_{-8.8}$ & $<26.92$ & $<27.13$ & \nodata & \nodata \\
Proxima Centauri & TIC 388857263 & $6.13^{+0.55}_{-0.55}$ & $26.69^{+0.1}_{-0.1}$ & $28.23^{+0.1}_{-0.1}$ & $26.8^{+0.02}_{-0.02}$ & $28.33^{+0.02}_{-0.02}$ \\
Alpha Centauri B & TIC 471011144 & $6.13^{+0.55}_{-0.55}$ & $27.06^{+0.36}_{-0.36}$ & $27.19^{+0.36}_{-0.36}$ & $27.52^{+0.02}_{-0.02}$ & $27.65^{+0.02}_{-0.02}$ \\
Sun & \nodata & $4.57^{+0.02}_{-0.02}$ & $26.77^{+0.72}_{-0.72}$ & $26.77^{+0.72}_{-0.72}$ & \nodata & \nodata \\

\hline
\end{tabular}
\tablecomments{Summary of star age and X-ray luminosity analyzed in this work. X-ray values from \citet{booth2017improved}$^\dagger$ and \citet{merloni2024srg}$^\star$}
\end{table*}


\subsection{\citet{booth2017improved} Sample}
\subsubsection{Stellar Ages} 

The majority of our sample comes from \citet{booth2017improved}. Fifteen of the \citet{booth2017improved} targets have ages measured using asteroseismology. Additionally, ten of these stars have ages estimated from comprehensive asteroseismic modeling of individual oscillation frequencies in literature \citep{Miglio2005A&A, silva2015ages, aguirre2017standing}. In five other cases, \citet{booth2017improved} estimated ages themselves using the BAyesian STellar Algorithm (BASTA) code \citep{SilvaAguirre2015MNRAS, aguirre2022bayesian} based on asteroesismic frequency detections from \citet{Chaplin2014ApJS} and stellar parameters from \citet{Buchhave2015ApJ}. Three additional stars had ages estimated from white dwarf companion dating---two from literature, and one with analysis performed by \citet{booth2017improved}. Finally, six individually selected stars had relatively well-known ages determined through various different methods, including isochrone fitting, association with a subpopulation of stars, and white dwarf chronometry, as described in \citet{booth2017improved}. 

\subsubsection{X-ray Luminosity} 
\citet{booth2017improved} used a combination of archival and newly-proposed observations to determine the stars' X-ray luminosity. The five stars with asteroseismic ages calculated by \citet{booth2017improved} using BASTA were selected for dedicated X-ray observations with \textit{XMM-Newton} and \textit{Chandra}. All other stars' X-ray data were from archival \textit{XMM-Newton} and \textit{Chandra} X-ray observations. For \textit{XMM-Newton}, a source was counted as detected if the number of source counts exceeded the number of counts from a pure background signal by at least $3\sigma$, where $\sigma$ was estimated as the square root of the number of expected background counts. In the case of \textit{Chandra}, background signal is typically very low. Therefore, they used full Poisson statistics and calculated the inverse percent function, where for a given expected number of background counts,  the number of counts in the source region had a probability of less than 0.3\% of occurring as a random fluctuation. The source was counted as detected if the number of source counts was this number or larger. Otherwise, the number was used as the upper limit on the X-ray counts of the source. 

To determine the X-ray flux of the sources, spectra were extracted using the relevant analysis tools for each telescope. These were then fitted with an optically thin thermal plasma model using the XSPEC \citep{arnaud2003x} fitting software. The two variables that were fitted were coronal temperature and emission measure. Redshift was fixed at zero. From the best fitting model for each object, the flux was calculated from 0.2-2 keV, as this is where weakly active cool stars display most of their X-ray emission \citep{booth2017improved}. 

For source regions that contained less than $\approx$90 counts, there were not enough data points to fit a spectrum accurately. In this case, \citet{booth2017improved} estimated X-ray flux was obtained using the WebPIMMS tool,\footnote{\url{https://heasarc.gsfc.nasa.gov/cgi-bin/Tools/w3pimms/w3pimms.pl}} inputting the mean count rate of the source region. A conversion factor from counts to flux was then applied for each instrument depending on the sensitivity at the time of observation. Conversion factors used by \citet{booth2017improved} are listed in Table~\ref{conversionfactors}.

To convert from X-ray flux to scaled luminosity, we use the approach demonstrated by \citet{schmitt2004nexxus}. Here, they found that when scaling X-ray luminosity with stellar surface area $4\pi R_{\star}$, with $R_{\star}$ being the stellar radius, then stars of all spectral types show the same spread of this quantity. For stars previously analyzed by \citet{booth2017improved}, we use the radius and distance values reported in their work. All other stars' radius and distance values were taken from the TESS Input Catalog (TIC, \citealt{Stassun2019AJ}).



\subsection{New Additions to the Sample}
\subsubsection{Stellar Ages} \label{sec:newages}
In addition to the \citeauthor{booth2017improved} sample, we perform a literature search for stars with asteroseismic measurements from TESS. We identify an additional four stars with a reliable age estimate, ranging from 2.4 to 6.2\,Gyr, and an available X-ray luminosity measurement from eROSITA (see Section~\ref{erosita}). 
Our newly-added targets are $\alpha$ Mensae (TIC 141810080, \citealt{chontos2021tess}), $\zeta$ Tucanae (TIC 425935521, \citealt{huber202220}),  $\pi$ Mensae (TIC 261136679, \citealt{huber202220}), and 88 Leonis (TIC 219030951,  \citealt{metcalfe2021magnetic}).






\subsubsection{eROSITA}\label{erosita}
We obtained estimates of the X-ray luminosity for these new stars (and also new estimates of the X-ray luminosity for six stars of the \citealt{booth2017improved} sample) from the eROSITA all-sky survey, which recently released X-ray source catalogs and sky maps with unparalleled depth. We use data from the first six months of the survey (eRASS1), which cataloged 930,000 sources in the Western Galactic Hemisphere within the most sensitive energy range of 0.2-2.3 keV. For a source to be counted as detected, the radius of the circular aperture adopted for photometry and the Poisson False Alarm Probability (FAP) were used. For a given background level, Poisson statistics was used to estimate the minimum number of photons within the aperture, so that the corresponding FAP was below a certain threshold, thus yielding an X-ray source detection to a given confidence level. 

We crossmatch eRASS1 sources with the \textit{XMM-Newton} and \textit{ROSAT} catalogs to check for consistency among the reported luminosity measurements. To convert the eROSITA detections into X-ray fluxes  (and ultimately luminosities) in the 0.2-2.3 keV band, we determine a conversion factor from eROSITA counts to flux with WebPIMMS. We use a conversion factor which assumes zero redshift, zero line of sight of hydrogen absorption, thermal emission of with $\log$T=6.5, and solar abundances \citep{johnstone2015coronal, Telleschi_2005}. Sources found in both the archival \textit{Chandra}/\textit{XMM-Newton} and eROSITA data sets were found to be consistent within $1\sigma$ of each other.

\begin{table}[t]
\fontsize{7}{10}\selectfont
\begin{center}
\caption{X-ray conversion factors}
\label{conversionfactors}
\begin{tabular}{lc}
\hline
Instrument              &  Conversion factor \\ &($erg\,s^{-1}\,cm^{-2}\,count^{-1}$)            \\ \hline
\textit{XMM-Newton} PN Medium Filter\tablenotemark{$^\dagger$}                  & $1.03\times10^{-12}$ \\
\textit{Chandra} ACIS-I No Grating (Cycle 16)\tablenotemark{$^\dagger$}   & $2.42\times10^{-11}$ \\
\textit{Chandra} ACIS-I No Grating (Cycle 12)\tablenotemark{$^\dagger$}   & $1.52\times10^{-11}$\\
\textit{Chandra} ACIS-I HET Grating (Cycle 10)\tablenotemark{$^\dagger$}  & $2.16\times10^{-10}$ \\
eROSITA \tablenotemark{$^\star$}                               & $8.63\times10^{-13}$        \\\hline

\end{tabular}
\end{center}
\tablenotetext{^\dagger}{From \citet{booth2017improved}. }
\tablenotetext{^\star}{From this work. }
\end{table}


\section{Analysis}\label{sec:analysis}

\label{methods}

Once we assembled our sample, we begin to investigate the relationship between X-ray luminosity and stellar age. Following \citet{booth2017improved}, we explored a variety of power law relationships between X-ray luminosity and age, which we carried out by fitting linear models to the logarithm of each dataset. We performed these fits using Markov Chain Monte Carlo (MCMC) methods with a differential evolution sampler \citep{terbraak}. This allows us to account for intrinsic scatter and heteroscedasticity in our fit. To this end, our model used 200 walkers for 2000 chains where the within-chain and between-chain variance was around 1, with an upper limit of 1.005, indicating a convergence of the chains. 

\begin{figure*}
   \centering
    \includegraphics[width=0.85\textwidth]{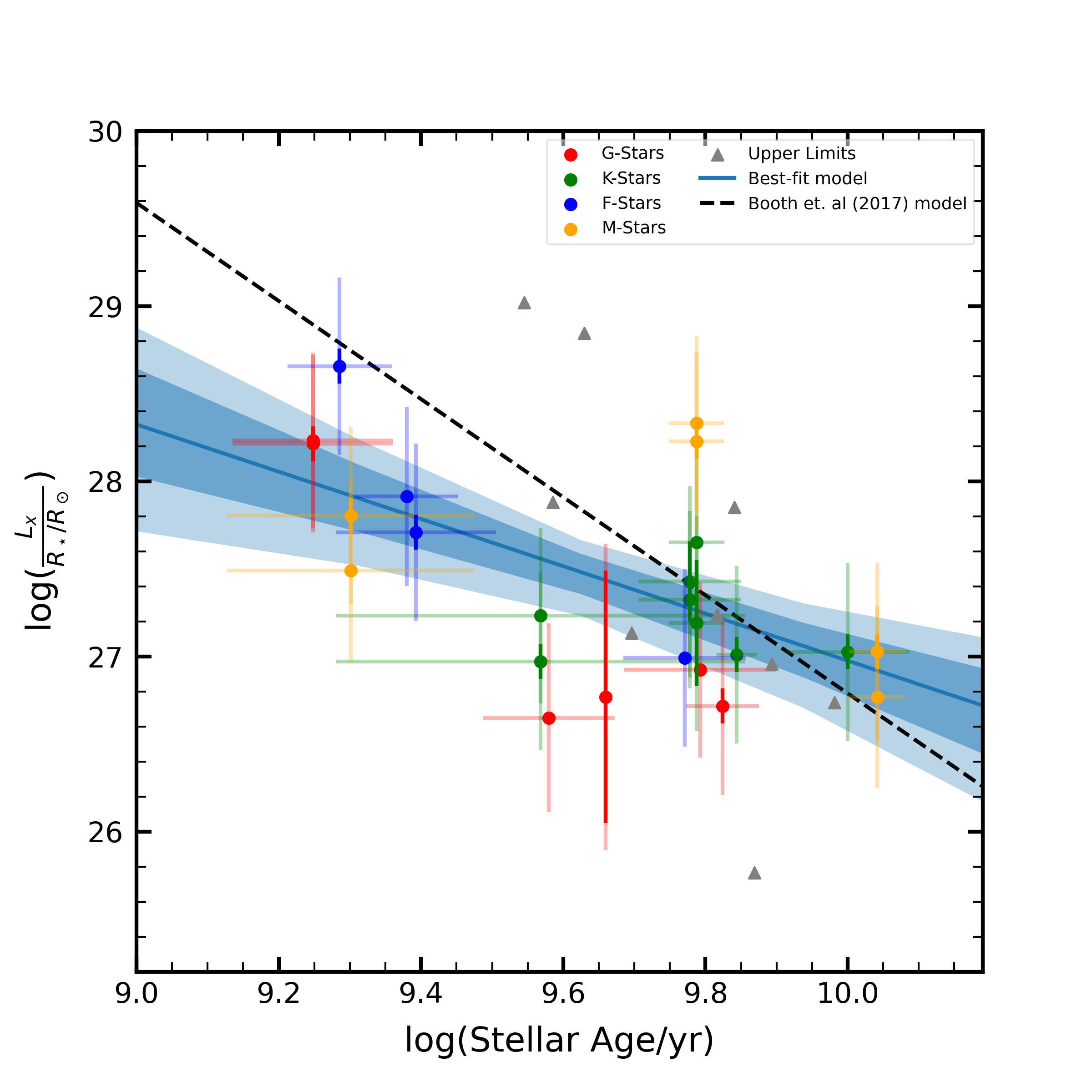}
   \caption{Relationship between X-ray luminosity and stellar age. We include both archival measurements from \citet{booth2017improved} and also incorporate newly measured targets with asteroseismic ages and X-ray luminosity from eROSITA. The points are color-coded by stellar type. The uncertainties are shown with a dark, smaller uncertainty reflecting the measurement uncertainties and the transparent larger uncertainties reflecting the addition of the jitter term. The dark and light blue shaded regions indicate the $1\sigma$ and $2\sigma$ uncertainties in the age-activity relationship, and the dark blue line indicates our best-fit model. Compared to the \citet{booth2017improved} model fit (shown by the dashed black line), we find a significantly shallower slope.}
   \label{fig:alltargets}
\end{figure*}

\subsection{Reproducing the \citet{booth2017improved} Analysis}\label{reproducingbooth}

We started our analysis by attempting to reproduce exactly the analysis performed by \citet{booth2017improved}. We implemented the same power law model they used as a linear model in logarithmic space:

\begin{equation} \label{eqn:boothmodel}
    \log{L_{x,n}} = m \log{t} + b
\end{equation}

\noindent where $\log L_{x,n}$ is the base 10 logarithm of X-ray luminosity normalized by the stellar surface area:
\begin{equation}
L_{x,n} = \frac{L_x}{R_\star/R_\odot}
\end{equation}

\noindent and $\log{t}$ is the base 10 logarithm of the stellar age, $m$ is the power law slope, and $b$ is the y-intercept in log space, which relates to the power law normalization constant. Both the values of $m$ and $b$ are parameters we fit in the MCMC. We assessed goodness of fit with a $\chi^2$ likelihood function: 

\begin{equation}
    \log\mathcal{L} =\sum_{i=1}^n \frac{-(\log L_{x,n,i} - \log L_{x,n,i,model})^2}{2 \sigma_i^2} +\log\sigma_i
\end{equation}

\noindent where $\log L_{x,i}$ is the measured value of the logarithm of scaled X-ray luminosity and $\log L_{x,i,model}$ are the parameters we fit in the MCMC. Because there are significant uncertainties in both the stellar luminosity and age, we accounted for both in our likelihood. Following \citet{press1992}, we defined the effective uncertainty, $\sigma_i$, as: 

\begin{equation}
    \sigma_i = \sqrt{\sigma_{Lx,n,i}^2 + (a\sigma_{t,i})^2}
\end{equation}

\noindent where $\sigma_{Lx,i}^2$ is the uncertainty in the logarithm of scaled X-ray luminosity and $\sigma_{t,i}$ is uncertainty in the logarithm of stellar age. 

We fit this linear model to the data using Differential Evolution Markov Chain Monte Carlo (DEMCMC, \citealt{terbraak}), as implemented by the {\tt edmcmc} package \citep{vanderburgedmcmc} and we estimate a power law slope of $-2.80 \pm 0.37$ and an intercept of $54.79\pm3.68$ (see Table~\ref{tab:slopes}). These values are consistent with \citet{booth2017improved}, where they estimate a power law slope value of $-2.80 \pm 0.72$ and an intercept of $54.65\pm6.98$. We note that the somewhat different uncertainties in the fit parameters are likely due to the fact that \citet{booth2017improved} used a different method, orthogonal distance regression, to fit the dataset. 

While this value is nearly identical to the one reported by \citet{booth2017improved}, we observed significant scatter around the best-fit model, with many sources having X-ray luminosities $\gg 1\sigma$ from the predicted value. We hypothesized that the error bars may be underestimated and decided to quantify this using a $\chi^2$ test. This test returns a p-value reflecting the probability that we would see such a poor fit by random chance, assuming the error bars were correctly estimated. We calculated a low p-value for our fit of $9\times10^{-7}$, indicating it is unlikely that such a poor fit would be the result of random chance. Instead, we think it is more likely that additional sources of scatter are present in the relationship. 

\begin{figure*}
   \centering
   \includegraphics[width=\textwidth]{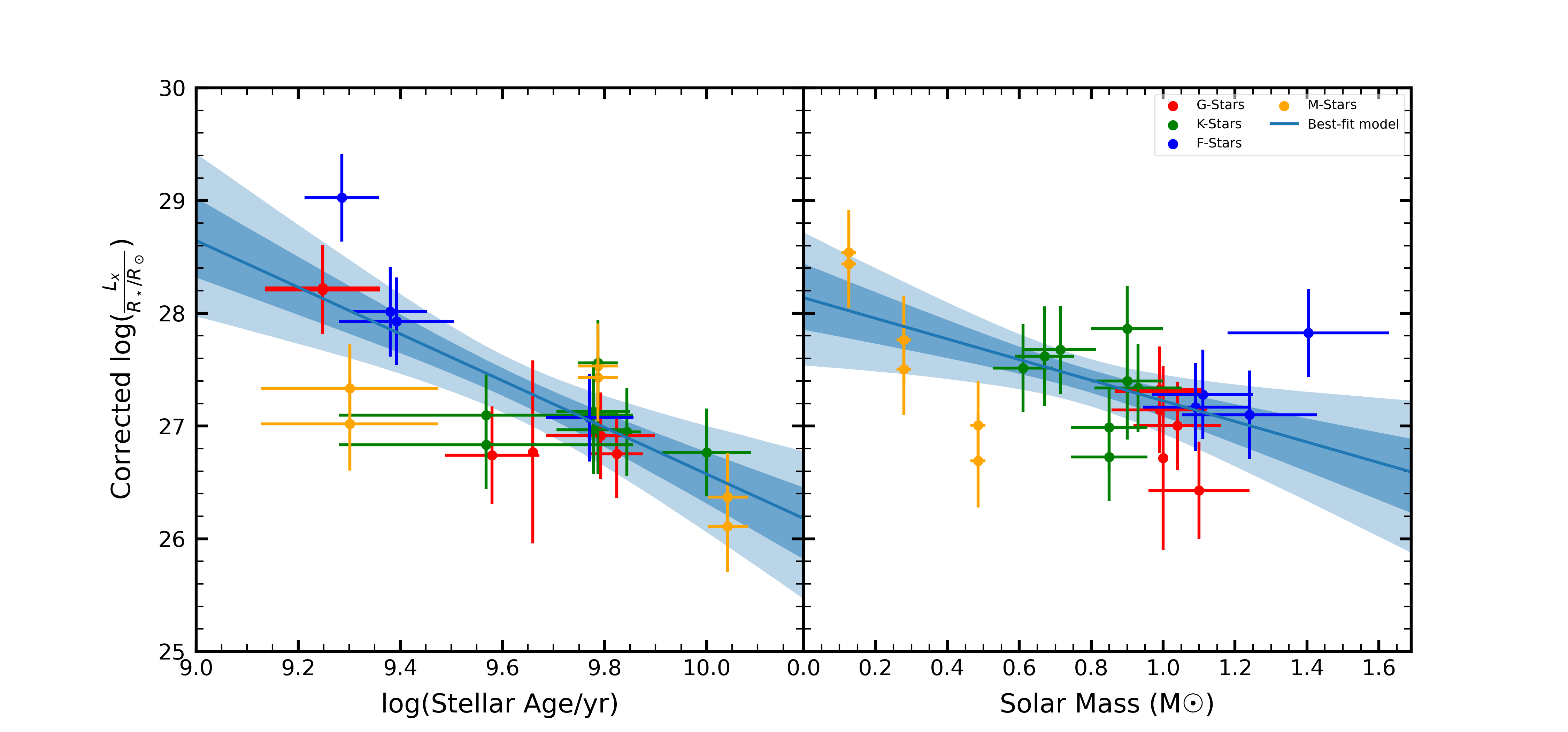}
   \caption{The X-ray luminosity/age relationship incorporating mass dependence. (a) Measurements of and best-fit relationship between X-ray luminosity as a function of stellar age, after subtracting the best-fit mass dependence.  (b)  Measurements of and best-fit relationship between X-ray luminosity as a function of stellar mass. As in Figure \ref{fig:alltargets}, the dark and light blue shaded regions indicate the $1\sigma$ and $2\sigma$ uncertainties in the age/mass/X-ray luminosity. We see tentative ($\approx 3\sigma$) evidence for a mass dependence in the X-ray luminosity/age relationship.}
   \label{fig:age_mass_relationship}
\end{figure*}



\subsection{Accounting for Astrophysical Variability} \label{jitter}
The results of our $\chi^2$ test performed in Section \ref{reproducingbooth} strongly suggest that the reported uncertainties in the X-ray luminosity are underestimated, or that variables other than stellar age impact a star's X-ray luminosity. This is not particularly surprising, because many processes have the potential to affect the observed X-ray luminosity, including cyclical variations of magnetic activity, rotational variations, and changing levels of circumstellar material \citep{neuhauser1995absorption}. To account for these effects, we include a jitter term in our model that is added in quadrature to our measurement uncertainties \citep{ford2006improving}. In particular, we modified our effective uncertainty so that it became: 

\begin{equation}
    \sigma_i = \sqrt{\sigma_{Lx,n,i}^2 + (a\sigma_{t,i})^2+\sigma_{j}^2}
\end{equation}

\noindent where $\sigma_j$ is the ``jitter'' parameter that we use to inflate our measurement uncertainties to account for astrophysical scatter. We fitted the X-ray luminosity/age relationship again with the addition of this new parameter, again including only the data used by \citet{booth2017improved}. When we include the ``jitter'' term, we find a slope of $-1.66 \pm 0.59$ (see Table~\ref{tab:slopes}). The quality of the fit is considerably improved compared to the model without a ``jitter'' term. When we add the best-fit jitter value in quadrature to our original uncertainties and perform the $\chi^2$ test again, we find a p-value of 0.94, indicating a much higher probability that the scatter observed could be due to random fluctuations given our inflated uncertainties.

Interestingly, we find that when we improve the qualify of the fit by including the ''jitter'' term, we measure a power law slope that is considerably shallower than what both we and \citet{booth2017improved} found without accounting for astrophysical scatter in the relationship. Unlike both of those analyses, the slope we measure is consistent with the slope of the age-activity relationship for younger stars ($-1.09 \pm 0.28$ to $-1.40 \pm 0.11$, \citealt{jackson2012coronal}) at the $1\sigma$ level. Evidently, the precise slope we measure, and the claim of a faster decline in X-ray luminosity at old ages is dependent on the noise model we choose and our assumptions about astrophysical variability. 


\subsection{Including New Stars and Measurements}
\label{newdata}

Next, we expanded our dataset to include new targets and observations to see whether adding new observations affects our conclusions about the slope of the age/X-ray luminosity relationship. In particular, we included in our sample the four new targets with asteroseismological ages (described in Section \ref{sec:newages}), as well as new eROSITA measurements of the X-ray luminosity for six of the targets included in the \citet{booth2017improved} sample (as described in Section \ref{erosita}). For targets with measurements of X-ray flux from both \citet{booth2017improved} and eROSITA, we included each X-ray luminosity measurement as an independent datapoint; this treatment has the effect of giving points with multiple measurements more weight because multiple observations taken years apart mitigates the effects of astrophysical variability. 

We performed the fit as described in Section \ref{jitter} with a ``jitter'' term included in the likelihood function to account for astrophysical variability in the X-ray luminosity. Using this combined dataset, we measure a slope of $-1.37\pm0.47$ (see Table \ref{tab:slopes}, Figure \ref{eqn:boothmodel}). This value is slightly shallower, but highly consistent with the slope we found from the original \citet{booth2017improved} when including jitter in our model. It is also consistent with the slopes reported for younger stars \citep{jackson2012coronal}. This reinforces the notion that there may not be a significant change in how quickly X-ray luminosity decreases for stars older than 1 Gyr. 

\subsection{Assessing the Impact of Upper Limits}
\label{upperlimits}

All of our fits thus far have only included stars with measured X-ray fluxes, ignoring upper limits from non-detections, as in \citet{booth2017improved}. While this approach is commonly taken, it could potentially bias the measured relationship by ignoring stars with the lowest X-ray luminosities. To determine whether this may effect our results, we recalculate the best-fit model when accounting for the available upper limits. Following \citet{Jenkins2024MNRAS}, we test two different probability distributions. First, we take these targets to have a mean $\log(L_X)$ of 0, with 3$\sigma$ uncertainties corresponding to their upper limits. Second, we assume a sigmoid probability distribution, where we include a fourth parameter in our fit that determines the steepness of the sigmoid slope. In both cases, we recover slopes ($-1.37\pm0.43$ and $-1.43\pm0.44$) and intercepts ($40.66\pm4.17$ and $41.20\pm4.33$) consistent with our previous result. We conclude that incorporating the upper limits does not appreciably change the results of our fits. 



\subsection{Testing for Mass Dependence in the X-ray Luminosity/Age Relationship}\label{massdependence}
Finally, we tested whether another variable --- stellar mass --- could be important in predicting the X-ray luminosity of old stars. Previous research found a relation between a star's decay of X-ray luminosity and their mass at ages younger than 1 Gyr \citep{preibisch2005evolution}. This due to the changes in how magnetic activity changes between high mass and low mass stars\citep{icsik2023scaling}.  \citet{booth2017improved} noted a tentative trend with mass in their results.  We tested explicitly whether the age/X-ray luminosity relationship has a mass dependence by explicitly including the a term for the stellar mass in our model: 
\begin{equation}
    \log{L_{x,n}} = m \log{t} + b + c * (M_{\star}-M_\odot)
\end{equation} 

Here, $M_{\star}$ is the stellar mass, $M_{\odot}$ is the mass of the Sun, and $c$ is a free parameter in our fit controlling the strength of the mass dependence. We fitted the model using MCMC as before, using the likelihood function including jitter described in Section \ref{jitter}. 

We find tentative evidence for mass dependence in the X-ray luminosity/age relationship for old stars; we measure a value of $c = -0.94 \pm 0.35$ for our mass dependence parameter. We find that this parameter is nonzero with almost $3\sigma$ confidence (p= $0.005$). Meanwhile, with the inclusion of mass dependence in the fit, we measure the power law slope of the age/activity relationship to be $-2.11\pm0.54$ (see Table \ref{tab:slopes}). Interestingly, including mass dependence in the X-ray age/luminosity relationship yields a somewhat steeper slope than our other fits including a jitter term to account for astrophysical scatter, but still consistent with our other slope measurements at the roughly $1\sigma$ level. This slope is also consistent with the slope measured by \citet{booth2017improved} at roughly the $1 \sigma$ level.  These results are therefore somewhat ambiguous; given the large uncertainties in the measured slopes, this test does not strongly favor either a shallow slope like that found for younger stars, or the steeper slope found by \citet{booth2017improved}. 


\begin{deluxetable*}{lcccc}
\tablecaption{Slope values\label{tab:slopes}}
\tablewidth{0pt}
\tablehead{
\colhead{Model Fit} & \colhead{Power-Law Slope} & \colhead{Intercept} & \colhead{Jitter} & \colhead{Mass Dependence}
}
\startdata
Replicating \citet{booth2017improved} (Section \ref{reproducingbooth}) & $-2.80 \pm 0.37$  & $54.79\pm3.68$ & \nodata  & \nodata\\
Modeling ``jitter'' (Section \ref{jitter})& $-1.66 \pm 0.59$ & $43.47\pm5.76$ & $0.46\pm0.16$  & \nodata \\
Modeling ``jitter'' and new data (Section \ref{newdata}) & $-1.37 \pm 0.47$ & $40.74\pm4.22$ & $0.48\pm0.094$  & \nodata \\
Modeling ``jitter'' and mass dependence (Section \ref{massdependence}) & $-2.11 \pm 0.54$ & $54.58\pm7.47$ & $0.3\pm0.20$  & $-1.06\pm0.48$\\
\\
\hline
Old stars from \citet{booth2017improved} & $-2.80 \pm 0.72$ & \nodata & \nodata  & \nodata \\
Young stars from \citet{jackson2012coronal} & $-1.09 \pm 0.28$ to $-1.40 \pm 0.11$ & \nodata & \nodata  & \nodata\\
\enddata
\tablecomments{Summary of age-activity slope values}
\end{deluxetable*}

\section{Discussion}\label{sec:discussion}

\subsection{Updated Understanding of the X-ray Luminosity/Age Relationship}

In our exploration, we drew several important conclusions that update our understanding of the relationship between X-ray luminosity and age for old stars. First, we found evidence for a considerable level of excess scatter ($\approx 0.5$ dex, see Table \ref{tab:slopes}) in the X-ray luminosity/age relationship. Evidently, other astrophysical processes or stellar properties can affect the X-ray luminosity of older main-sequence stars besides just age. This is not surprising, but must be accounted for when modeling the relationship. 

When we modeled the excess scatter in the age/X-ray luminosity relationship, we found that the power law slope was shallower than previously reported. \citet{booth2017improved} measured a power law slope of $-2.8$, but when we include a ``jitter'' term in our model to account for excess scatter due to astrophysical variations, we measure slopes between $-1.37$ and $-2.11$ on largely the same dataset. These slope values are consistent with the power law slopes calculated for young stars (less than 1 Gyr in age), indicating that there might not be a break in the stellar spin-down and activity decrease relation between old and young stars.

Finally, we detected tentative (almost $3\sigma$) evidence for mass dependence in the X-ray luminosity/age relationship. This is not unexpected --- M-type dwarf stars have very different internal structures and magnetic field properties compared to Sun-like G-type dwarf stars, so it stands to reason they would have different X-ray luminosities as well. Given the small sample size we have at our disposal (28 stars) and the relatively high level of excess scatter we observe in the relationship, considerably more stars with both well-determined ages and X-ray luminosities are needed to solidify this detection. 


 


\subsection{Prospects for Using X-ray Luminosity to Measure Stellar Ages} 

X-ray luminosity has been seen as a promising method to determine the ages of main sequence stars older than about 1 Gyr. After a turbulent youth showcasing rapid contraction onto the zero-age main sequence and high levels of magnetic activity, mature stars (older than about 1 Gyr) change very slowly as they burn hydrogen on the main sequence. Many of our most widely-applicable age measurement techniques, like isochronal age-dating, lithium depletion, and cluster dating become more challenging or impossible at these ages. If X-ray luminosity could be calibrated to give a precise age estimate for older main sequence stars, it would be a very powerful tool. 

Unfortunately, two of our main results pose challenges to using X-ray luminosity as an age proxy. The presence of astrophysical scatter in the relationship sets a maximum precision with which age can be estimated for an individual star because it becomes impossible to distinguish whether an unusually low X-ray flux indicates old age or some other process/property. Making matters worse, we also found that the slope of the age/activity relationship is likely shallower than previously believed \citep{booth2017improved}. If X-ray luminosity does decrease at this slower rate than previously believed, it makes changes in luminosity due to age less significant compared to the astrophysical scatter, further impeding its use for age estimation.  

One possible way forward to mitigate these challenges would be to average multiple X-ray luminosity measurements for each star. We often saw significant (much greater than the measurement uncertainties) differences between X-ray fluxes measured for the same star at different times by Chandra/XMM Newton and eROSITA. We attribute this to time variability in the X-ray luminosity due to processes like stellar rotation or magnetic activity cycles. Increasing the number of X-ray observations of each star could trace out these changes and make it possible to estimate a time-averaged X-ray luminosity for each stars. 

Another way would be to investigate other predictors/correlates. We have already detected a tentative correlation between stellar mass and X-ray luminosity, and accounting for this reduces the scatter in the X-ray luminosity/age relationship. It would be interesting to investigate whether other stellar properties like metallicity, inclination angle, or proximity to stellar/planetary companions also correlate with X-ray luminosity and can be used to tighten the relationship. 

If we can combine multiple observations over year--decade timescales to measure a time-averaged X-ray luminosity, and identify other variables that correlate with X-ray luminosity, it is plausible that much of the excess scatter we find in the age/X-ray luminosity relationship can be removed, making X-ray luminosity again a promising way to estimate stellar ages. Accomplishing this would require a concerted observational effort, but it might be worthwhile in order to realize the goal of an effective clock to measure stellar ages throughout the main-sequence phase. Until then,  X-ray luminosity cannot by itself precisely constrain a star's age, but could be used as part of a broader ensemble of age measurements.



\section{Summary}\label{sec:summary}
In this work, we present X-ray detections of 28 stars with determined ages above 1 Gyr. We use data previously analyzed by \citet{booth2017improved}, and include targets found in the newly released eROSITA catalogue to reassess this relationship. The main conclusions of this work are: 
\begin{itemize}
\item X-ray luminosity decreases with age more slowly than previously thought. We measure power law slopes between $-1.37$ and $-2.11$ for the decrease of X-ray luminosity over time, compared to a slope of -2.8 from \citet{booth2017improved}.
\item We find evidence for astrophysical scatter in the X-ray luminosity/age relationship. When we allow a ``jitter'' term to model excess scatter in the relationship, we find typical values of about 0.5 dex, or about a factor of 3. 
\item We find tentative ($\approx 3 \sigma$) evidence for mass dependence in the X-ray luminosity/age relationship. 
\item The shallower slope and astrophysical scatter in the X-ray/age relationship pose challenges for using X-ray luminosity as an age proxy, but identifying other parameters (like mass) that can help predict X-ray luminosity might help tighten the relationship.  

\end{itemize}
We note that a larger sample size is needed to better constrain the relationship between stellar X-ray luminosity and age. While X-ray luminosity alone cannot provide precise stellar age measurements, it may be used in conjunction with other age-dating techniques.

\acknowledgments
This material is based on work supported by the MIT Summer Research Program (MSRP). 
SAJ was funded in part by the National Science Foundation Graduate Research Fellowship under Grant No. 1745302.
MSF gratefully acknowledges the generous support provided by NASA through Hubble Fellowship grant HST-HF2-51493.001-A awarded by the Space Telescope Science Institute, which is operated by the Association of Universities for Research in Astronomy, Inc., for NASA, under the contract NAS 5-26555.

This work is based on data from eROSITA, the soft X-ray instrument on board SRG, a joint Russian-German science mission supported by the Russian Space Agency (Roskosmos), in the interests of the Russian Academy of Sciences represented by its Space Research Institute (IKI), and the Deutsche Zentrum für Luft- und Raumfahrt (DLR). The SRG spacecraft was built by Lavochkin Association (NPOL) and its subcontractors and is operated by NPOL with support from the Max Planck Institute for Extraterrestrial Physics (MPE). The development and construction of the eROSITA X-ray instrument was led by MPE, with contributions from the Dr.~Karl Remeis Observatory Bamberg \& ECAP (FAU Erlangen-Nuernberg), the University of Hamburg Observatory, the Leibniz Institute for Astrophysics Potsdam (AIP), and the Institute for Astronomy and Astrophysics of the University of Tübingen, with the support of DLR and the Max Planck Society. The Argelander Institute for Astronomy of the University of Bonn and the Ludwig Maximilians Universität Munich also participated in the science preparation for eROSITA.
This work has used data from the European Space Agency (ESA) mission \emph{Gaia},\footnote{\url{https://www.cosmos.esa.int/gaia}} processed by \emph{Gaia} Data Processing and Analysis Consortium (DPAC).\footnote{\url{https://www.cosmos.esa.int/web/gaia/dpac/consortium}} 
This research has used the VizieR catalog access tool, CDS, Strasbourg, France. The original description of the VizieR service was published in A\&AS 143, 23. 

\facilities{\erosita, \Gaia{} \citep{GaiaDR3}, Mikulski Archive for Space Telescopes \citep{MAST}}

\software{ \texttt{astroquery} \citep{astroquery}, \texttt{corner.py} \citep{corner}, \texttt{edmcmc} \citep{vanderburgedmcmc}, \texttt{matplotlib} \citep{plt}, \texttt{numpy} \citep{np}}

\clearpage

\bibliographystyle{aasjournalmod}
\bibliography{bibliography.bib}


\end{document}